\documentclass[aps,showpacs,superscriptaddress,preprintnumbers,amsmath,amssymb]{revtex4}
\usepackage{bm,graphicx}

\begin{document}

\preprint{Submitted to Eur. Phys. J. E}

\title{Elastic interaction between ``hard'' or ``soft'' pointwise
inclusions\\ on biological membranes}

\author{Denis Bartolo}
\affiliation{Laboratoire de Physico-Chimie Th\'eorique, ESPCI, 10 rue
Vauquelin, F-75231 Paris cedex 05, France}

\author{Jean-Baptiste Fournier}
\affiliation{Laboratoire de Physico-Chimie Th\'eorique,
ESPCI, 10 rue Vauquelin, F-75231 Paris cedex 05,
France}
\affiliation{F\'ed\'eration de Recherche~CNRS~2438 ``Mati\`ere et Syst\`emes
Complexes"}

\begin{abstract}
We calculate the induced elastic-interaction between pointwise membrane
inclusions that locally interact up to quadratic order with the membrane
curvature tensor. For isotropic inclusions, we recover the usual
interaction proportional to the inverse fourth power of the separation,
however with a prefactor showing a non-trivial dependence on the
rigidity $\Gamma$ of the quadratic potential. In the large $\Gamma$
limit, corresponding to ``hard'' inclusions, we recover the standard
prefactor first obtained by Goulian et al. [Europhys.  Lett.
\textbf{22}, 145 (1993)]. In the small $\Gamma$ limit, corresponding to
``soft'' inclusions, we recover the recent result of Marchenko and
Misbah [Eur. Phys. J. E \textbf{8}, 477 (2002)]. This shows that the
latter result bears no fundamental discrepancy with previous works,
but simply corresponds to the limit of soft inclusions. We discuss
how the same inclusion can be depicted as hard or soft according to the
degree of coarse-graining of the pointwise description. Finally, we
argue that conical transmembrane proteins should be fundamentally
considered as hard inclusions.
\end{abstract}

\date{\today}
\pacs{87.16.Dg (Membranes, bilayers, and vesicles), 87.15.Kg (Molecular
interactions; membrane-protein interactions)}
\maketitle
\section{Introduction}
In a recent paper~\cite{misbah02}, Marchenko and Misbah have revisited
the calculation of the long-range, membrane-mediated interaction acting
between inclusions embedded in biological or surfactant membranes. Using
a method inspired from electrostatics, they calculated an
``induced interaction'', which they compared with the results of several
works previously published on the
subject~\cite{goulian93,park96,kim98,dommersnes99,dommersnes02}. They
pointed out two major discrepancies: (i) their result is
proportional to~$1/\kappa^2$ ($\kappa$ being the bending rigidity of the
membrane) while the induced interaction calculated in
Refs.~\cite{goulian93,park96,kim98,dommersnes99,dommersnes02} is
directly proportional to $\kappa$, (ii) the result of
Refs.~\cite{goulian93,park96,kim98,dommersnes99,dommersnes02} involves
the size of the inclusion (either directly or through a cutoff in the
case of pointwise descriptions) whereas the size of the inclusion does
not appear in the result of Ref.~\cite{misbah02}.

In this contribution, we show that there are no discrepancies between
the approaches of
Refs.~\cite{goulian93,park96,kim98,dommersnes99,dommersnes02} and that
of Ref.~\cite{misbah02}, neither in the spirit of the model nor in the
results.  We demonstrate that the different expressions obtained for the
induced interaction correspond to different limits of a common, more
general model. We do so by solving a pointwise quadratic
interaction model similar to the one used in Ref.~\cite{misbah02}, exactly,
instead of perturbatively, and without the need to introduce a
high-wavector cutoff. According to the rigidity $\Gamma$ of the
quadratic interaction potential, we find either the standard prefactor first
obtained by Goulian et
al.~\cite{goulian93,park96,kim98,dommersnes99,dommersnes02}, in the large
$\Gamma$ limit corresponding to ``hard'' inclusions, or the result of
Marchenko and Misbah~\cite{misbah02}, in the small $\Gamma$ limit
corresponding to ``soft'' inclusions.  

In the second part of the paper we discuss the role of the cutoff
(introduced e.g. in Refs.~\cite{dommersnes99,dommersnes02}) in relation
with a coarse-graining procedure of the pointwise description. We
determine how the effective quadratic interaction potential depends on
the coarse-graining scale, and we discuss the distinction between hard
and soft inclusions in the light of this analysis.

\section{Induced interaction}\label{induced}

Here, we calculate the induced interaction between pointwise
``defects'', or inclusions. We describe the inclusions using the same
formalism  as in Ref.~\cite{misbah02}. We use, however, a more general
Hamiltonian to describe the elasticity of the membrane, and, in
addition, we perform our calculations exactly, whereas the result of
Ref.~\cite{misbah02} was obtained at lowest-order in a perturbative
expansion. We emphasize that we shall not introduce any cutoff during
our calculation.

The generalized membrane elastic Hamiltonian~\cite{helfrich73} we use
is
\begin{equation}
F_\mathrm{m}=\frac{1}{2}\int\!dx\,dy\,\left[
\gamma\,h^2+\sigma\left(\nabla h\right)^2
+\kappa\left(\nabla^2 h\right)^2
+\kappa_6\left(\nabla\,\nabla^2 h\right)^2
+\kappa_8\left(\nabla^4 h\right)^2
\right],
\end{equation}
where $h(x,y)$ describes the normal displacement of the membrane. Being
quadratic in $h(x,y)$, this Hamiltonian is accurate only for small membrane
deformations.  We have assumed that the membrane has no physical
boundary except at infinity, which allowed us to discard all the terms
that could be written as the divergence of a vector field (e.g., the
Gaussian curvature energy~\cite{helfrich73}). The parameters $\gamma$
and $\sigma$ describe a confining potential and a membrane tension,
respectively. In most cases they are negligibly small,
and they play no role except at very large scales. The
parameter $\kappa$ is the usual bending constant~\cite{helfrich73} and
$\kappa_6$, $\kappa_8$ are higher-order bending elasticity
constants~\cite{helfrich96}. We may rewrite the membrane Hamiltonian in
the reciprocal space as 
\begin{equation}\label{Fq}
F_\mathrm{m}=\int\!\frac{d^2q}{(2\pi)^2}\,
\frac{\kappa}{2}\left[
\ell_\gamma^{-4}+\ell_\sigma^{-2}\,q^2
+q^4\left(
1+b_6^2\,q^2+b_8^4\,q^4
\right)
\right]|h(\mathbf{q})|^2.
\end{equation}
Here $L$ is the lateral extension of the membrane;
$b_6=(\kappa_6/\kappa)^{1/2}$, $b_8=(\kappa_8/\kappa)^{1/4}$ are
nanometric lengths of the order of the thickness of the membrane, and
$\ell_\gamma=(\kappa/\gamma)^{1/4}$, $\ell_\sigma=(\kappa/\sigma)^{1/2}$
are very large lengths in the case of free membranes with a vanishing
tension.

To describe the coupling of the membrane with a pointwise isotropic
inclusion labeled $i$, situated in $\mathbf{r}=\mathbf{r}_i$, we generalize
the expression used in Ref.~\cite{misbah02} (in which the term
proportional to $\bar D$ below was omitted):
\begin{equation}\label{int}
U_i=\int\!dx\,dy\,\,\delta(\mathbf{r}-\mathbf{r}_i)
\left[
A\,\nabla^2h(\mathbf{r})-\frac{D}{2}\,
\partial_\alpha\partial_\beta h(\mathbf{r})\,
\partial_\alpha\partial_\beta h(\mathbf{r})
-\frac{\bar D}{2}\,
\partial_\alpha\partial_\alpha h(\mathbf{r})\,
\partial_\beta\partial_\beta h(\mathbf{r})
\right],
\end{equation}
where $\alpha,\beta\in\{1,2\}$ stand for the $x$ and $y$ coordinates and
$\delta(\mathbf{r})$ is the Dirac distribution. Summation over repeated
indices will be implied throughout. For a membrane containing two
isotropic inclusions, the total energy functional is
$F_\mathrm{m}+U_1+U_2$. To proceed, we rewrite the interaction terms
$U_i$, up to a constant, as
\begin{equation}\label{intcurv}
U_i=\frac{\Gamma}{2}\left(H^{(i)}_{\alpha\beta}
-C^{(i)}_{\alpha\beta}\right)^2+\frac{\Gamma\epsilon}{2}
H^{(i)}_{\alpha\alpha}H^{(i)}_{\beta\beta},
\end{equation}
where $H^{(i)}_{\alpha\beta}\equiv\partial_\alpha\partial_\beta
h(\mathbf{r}_i)$, and $C^{(i)}_{\alpha\beta}=c_0\,\delta_{\alpha\beta}$
for an isotropic inclusion. The constants $\Gamma$, $\epsilon$ and
$c_0$ are given by $\Gamma=-D$, $\epsilon=\bar D/D$ and  $c_0=A/D$. Note
that the condition $\epsilon>-\frac{1}{2}$ has to be satisfied, overwise
the membrane would be locally unstable. In order to give a simple
interpretation of the interaction, we can also write Eq.~(\ref{intcurv})
in term of the principal values $H_1^{(i)}$ and $H_2^{(i)}$ of the local
curvature tensor $H^{(i)}_{\alpha\beta}$, i.e.,
\begin{equation}
U_i=\frac{1}{2}(1+2\epsilon)\,\Gamma\left[\left(H_1^{(i)}-c\right)^2+
\left(H_2^{(i)}-c\right)^2\right]-
\frac{1}{2}\Gamma\epsilon\left(H_1^{(i)}-H_2^{(i)}\right)^2.
\end{equation}
The first two terms show that the inclusion locally promote an isotropic
curvature $c$, and the third term distinguishes conformations for which the
deviation of $H_1^{(i)}$ and $H_2^{(i)}$ from $c$ have identical or
opposite signs. Note that it has been implicitely assumed that
$\Gamma>0$.  

In order to calculate the interaction energy between the two inclusions,
we must minimize $F_\mathrm{m}+U_1+U_2$ with respect to the shape of the
membrane. We do this in two steps: (i) first, we minimize $F_\mathrm{m}$
with respect to the membrane shape while setting prescribed values of
$H^{(i)}_{\alpha\beta}$, which yields
$F_\mathrm{min}(H^{(1)}_{\alpha\beta},H^{(2)}_{\alpha\beta})$, (ii)
next, we minimize
$F_\mathrm{tot}=F_\mathrm{min}(H^{(1)}_{\alpha\beta},H^{(2)}_{\alpha\beta})+U_1+U_2$
with respect to $H^{(1)}_{\alpha\beta}$ and $H^{(2)}_{\alpha\beta}$.

The first step was performed in Ref.~\cite{dommersnes02}. The result is
\begin{equation}\label{res1}
F_\mathrm{min}\left(H^{(1)}_{\alpha\beta},H^{(2)}_{\alpha\beta}\right)=
\frac{1}{2}\kappa\,\mathsf{K^t}\,\mathsf{M}^{-1}\,\mathsf{K}.
\end{equation}
In this expression, the subscript $\mathsf{t}$ means transpose,
\begin{equation}
\mathsf{K^t}=(H^{(1)}_{11},H^{(1)}_{12},H^{(1)}_{22},
H^{(2)}_{11},H^{(2)}_{12},H^{(2)}_{22}),
\end{equation}
$\mathsf{M}$ is the $6\times6$ matrix:
\begin{equation}
\mathsf{M}=
\begin{pmatrix}
G_{1111}(0)&G_{1112}(0)&G_{1122}(0)&
G_{1111}(\mathbf{r})&G_{1112}(\mathbf{r})&G_{1122}(\mathbf{r})\cr
G_{1112}(0)&G_{1122}(0)&G_{1222}(0)&
G_{1112}(\mathbf{r})&G_{1122}(\mathbf{r})&G_{1222}(\mathbf{r})\cr
G_{1122}(0)&G_{1222}(0)&G_{2222}(0)&
G_{1122}(\mathbf{r})&G_{1222}(\mathbf{r})&G_{2222}(\mathbf{r})\cr
G_{1111}(-\mathbf{r})&G_{1112}(-\mathbf{r})&G_{1122}(-\mathbf{r})&
G_{1111}(0)&G_{1112}(0)&G_{1122}(0)\cr
G_{1112}(-\mathbf{r})&G_{1122}(-\mathbf{r})&G_{1222}(-\mathbf{r})&
G_{1112}(0)&G_{1122}(0)&G_{1222}(0)\cr
G_{1122}(-\mathbf{r})&G_{1222}(-\mathbf{r})&G_{2222}(-\mathbf{r})&
G_{1122}(0)&G_{1222}(0)&G_{2222}(0)\cr
\end{pmatrix}
\end{equation}
in which $\mathbf{r}=\mathbf{r}_1-\mathbf{r}_2$, and
$G_{\alpha\beta\gamma\delta}(\mathbf{r})=\partial_\alpha\partial_\beta
\partial_\gamma\partial_\delta G(\mathbf{r})$, where
\begin{equation}\label{Green}
G(\mathbf{r})=\int\!\frac{d^2q}{(2\pi)^2}\,
\frac{e^{i\mathbf{q}\cdot\mathbf{r}}}
{\ell_\gamma^{-4}+\ell_\sigma^{-2}\,q^2
+q^4\left(
1+b_6^2\,q^2+b_8^4\,q^4
\right)}
\end{equation}
is the Green function associated with the elastic Hamiltonian. Note that
no cutoff is required for the convergence of
$G_{\alpha\beta\gamma\delta}(0)$ thanks to terms of order $q^6$ and
$q^8$. The question of the high-wavevector cutoff will be raised
later.

To perform the second step of the minimization, we rewrite the
interaction term as
\begin{equation}\label{u1plusu2}
U_1+U_2=\frac{\Gamma}{2}\left(\mathsf{K}-\mathsf{C}\right)^\mathsf{t}
\mathsf{N}
\left(\mathsf{K}-\mathsf{C}\right),
\end{equation}
where
\begin{equation}
\mathsf{C^t}=(C^{(1)}_{11},C^{(1)}_{12},C^{(1)}_{22},
C^{(2)}_{11},C^{(2)}_{12},C^{(2)}_{22}),
\end{equation}
contains the curvatures attempted to be set by the inclusions, and N is
the $6\times6$ block-diagonal matrix: 
\begin{equation}\label{Nmatrix}
\mathsf N=
\begin{pmatrix}
1+\epsilon&0&\epsilon\cr
0&2\epsilon&0\cr
\epsilon&0&1+\epsilon\cr
\end{pmatrix}
\otimes
\begin{pmatrix}
1&0\cr
0&1\cr
\end{pmatrix},
\end{equation}
where $\otimes$ stands for the matrix direct 
product.

The resulting total energy $F_\mathrm {tot}=F_\mathrm {min}+U_1+U_2$
is quadratic in $\mathsf{K}$, and its minimization yields
\begin{equation}
F_\mathrm{tot,\,min}=-\frac{\Gamma^2}{2}\,
\mathsf{C^t}\mathsf{N}\left[
\kappa\,\mathsf{M}^{-1}+\Gamma\,\mathsf{N}
\right]^{-1}\mathsf{N}\mathsf{C}+\frac{\Gamma}{2}\,
\mathsf{C^t}\mathsf{N}\mathsf{C}.
\end{equation}
After some algebra, this expression can be recast in the simpler form,
similar to that of Eq.~(\ref{res1}):
\begin{equation}\label{centralresult}
F_\mathrm{tot,\,min}=\frac{1}{2}\kappa\,\mathsf{C^t}
\left(
\mathsf{M}+\frac{\kappa}{\Gamma}\,\mathsf{N}^{-1}
\right)^{-1}\mathsf{C}.
\end{equation}
Equation~(\ref{centralresult}) is very general and corresponds to our
central result: it describes the interaction between two pointwise
inclusions to all orders in their inverse separation $1/r$ (as long as
the Hamiltonian $F_m$ is fit to describe the membrane distortion energy at the
scale $r$), for any value of the rigidities $\Gamma$ and
$\Gamma\epsilon$, and actually for
isotropic or anisotropic inclusions, since the above derivation does not
assume any form for the components of $\mathsf{C}$. For anisotropic
inclusions, $\mathsf{C}$ should contain the components of the curvature tensor
favored by the inclusions~\cite{dommersnes02}.

To proceed further, we must first determine the components of the matrix
$\mathsf{M}$.  In the large $r$ limit, i.e., $r\gg b_6,b_8$ (recall that
$b_6$ and $b_8$ are nanoscopic lengths), and assuming also
$r\ll\ell_\gamma,\ell_\sigma$ (i.e., taking $\ell_\gamma\to\infty$ and
setting strictly $\ell_\sigma^{-1}=0$ for the sake of simplicity),
Eq.~(\ref{Green}) yields
\begin{equation}
G(\mathbf{r})\simeq C_1+\frac{1}{16\pi}r^2\ln(\frac{r^2}{C_2}).
\end{equation}
The constants $C_1$ and $C_2$ depend on $\ell_\gamma$, but they actually
disappear in the calculation of
$G_{\alpha\beta\gamma\delta}(\mathbf{r})$, leaving
$1/(16\pi)(x^2+y^2)\ln(x^2+y^2)$ as the only relevant contribution in
$G(\mathbf{r})$. To calculate the components of the matrix $\mathsf{M}$
in the large $r$ limit defined above, we set $x=r$ and $y=0$, yielding,
$G_{1111}=G_{1122}\simeq-1/(4\pi r^2)$, $G_{2222}\simeq3/(4\pi r^2)$,
and $G_{1112}=G_{1222}=0$. Next, we have
\begin{equation}\label{b2}
G_{\alpha\beta\gamma\delta}(0)=n_{\alpha\beta\gamma\delta}\times
\int\!\frac{dq}{16\pi}\,\frac{q^5}
{\ell_\gamma^{-4}+\ell_\sigma^{-2}\,q^2
+q^4\left(
1+b_6^2\,q^2+b_8^4\,q^4
\right)}\equiv b^{-2}n_{\alpha\beta\gamma\delta},
\end{equation}
where $n_{\alpha\beta\gamma\delta}$ contains the angular part of the
integration on $\mathbf{q}$. Simple calculations yield
$n_{1111}=n_{2222}=3$, $n_{1122}=1$, and $n_{1112}=n_{1222}=0$. For
$b_6\approx b_8\ll\ell_\gamma,\ell_\sigma$, as previously specified, the
constant $b$ is a nanometric length comparable to $b_6$ or $b_8$.
Therefore, without loss of generality, we have (still in the large $r$
limit):
\begin{equation}
\mathsf{M}\simeq\frac{1}{b^2}
\begin{pmatrix}
3&0&1\cr
0&1&0\cr
1&0&3\cr
\end{pmatrix}
\otimes
\begin{pmatrix}
1&0\cr
0&1\cr
\end{pmatrix}
+
\frac{1}{4\pi r^2}
\begin{pmatrix}
-1&0&1\cr
0&-1&0\cr
-1&0&3\cr
\end{pmatrix}
\otimes
\begin{pmatrix}
0&1\cr
1&0\cr
\end{pmatrix}
\end{equation}
To calculate the interaction $F_\mathrm{int}(r)$ between two identical
isotropic inclusions, we set 
\begin{equation}
\mathsf{C^t}=(c,0,c,c,0,c),
\end{equation}
and, using Eq.~(\ref{centralresult}), we obtain:
\begin{equation}\label{fint}
F_\mathrm{int}(r)=\frac{\kappa\Gamma^3 c^2(1+2\epsilon)^2}
{2\pi^2
 \left(\kappa+2\Gamma b^{-2}\right)
\left[\kappa+4\Gamma b^{-2}\left(1+2\epsilon\right)\right]^2}
\,\frac{1}{r^4}+\mathcal{O}\left(\frac{1}{r^6}\right).
\end{equation}

We now consider two limits: \textit{hard} inclusions, corresponding to
$\Gamma/\kappa\gg b^2$, and \textit{soft} inclusions, corresponding
to $\Gamma/\kappa\ll b^2$ (note that we keep $\epsilon$ of order unity for the
sake of simplicity). In the former case, the inclusions
strongly set their preferred curvature; in the latter case, they
only weakly tend to establish it.

For \textit{hard} inclusions, we let $\Gamma$ go to infinity, and to
leading order in $1/r$, we obtain 
\begin{equation}\label{hard}
F_\mathrm{int,\,hard}(r)\simeq
\frac{\kappa\,b^2
c^2}{64\pi^2}\,\left(\frac{b}{r}\right)^4.
\end{equation}
We thus recover the form obtained in
Refs.~\cite{goulian93,park96,kim98,dommersnes99,dommersnes02}, with
$\kappa$ appearing in the numerator. Here the microscopic length $b$
arising from the membrane Hamiltonian plays the role of the cutoff
introduced in Refs.~\cite{park96,kim98,dommersnes99,dommersnes02}, or
the role of the radius of the disk modeling the inclusion in the case of
Ref.~\cite{goulian93}. 

For \textit{soft} inclusions, we obtain to leading order in $1/r$ and at
lowest-order in $\Gamma$ (like in Ref.~\cite{misbah02}):
\begin{equation}\label{soft}
F_\mathrm{int,\,soft}(r)\simeq
\frac{\Gamma^3\,c^2}{2\pi^2\kappa^2\,r^4}(1+2\epsilon)\equiv
-\frac{2A^2D}{4\pi^2\kappa^2\,r^4}(1+2\epsilon),
\end{equation}
where $\kappa$ appears now to the second power in the denominator. This
expression coincides exactly with the Eq.~(37) of Ref.~\cite{misbah02},
in which $\epsilon=0$ was implicitely assumed. Contrary to the case of
hard inclusions, the microscopic length $b$ does not appear in the
interaction, it would however reappear in the next order term
$\mathcal{O}(\Gamma^4)$. We recall that $1+2\epsilon$ is positive, hence
the interaction is always repulsive in the present case of isotropic
inclusions.

We have therefore shown that there is no fundamental discrepancy
between the recent result of Marchenko and Misbah~\cite{misbah02} and
the earlier results published in
Refs.~\cite{goulian93,park96,kim98,dommersnes99,dommersnes02}.  The
result of Ref.~\cite{misbah02} applies to soft inclusions, i.e., to
inclusions that weakly tend to curve the membrane, while the result of
Refs.~\cite{goulian93,park96,kim98,dommersnes99,dommersnes02} is valid
for hard inclusions that strongly enforce their preferred membrane
curvature.

\section{The role of the cut-off}

Let us now comment on the role of the high-wavevector cutoff. Of
course, any well-defined elastic theory does involve a coarse-graining
cutoff $a^{-1}$, and the elastic constants $\kappa(a)$,
$\kappa_6(a)$, and $\kappa_8(a)$ depend on it. Thanks to the high-order
bending elastic constants $\kappa_6$ and $\kappa_8$, there was no need
to introduce a cutoff to calculate the interaction between the
quadratic pointwise inclusions considered here.  This is however
somewhat arbitrary, since the microscopic length $b$ arising from this
procedure is expected to compare with the natural cutoff $a$. This
simply means that reintrocuding the cutoff would merely redefine $b$,
leaving its order of magnitude unchanged. It is therefore unimportant to
specify whether the cutoff was introduced or not in the procedure
defining the microscopic length $b$ that appears in the interaction. 

\section{Coarse-Graining of the membrane--inclusion system}

Describing an inclusion by a pointwise potential is of course a
mathematical approximation hiding some physical microscopic
length-scale.  One may actually choose the ``resolution'' of the
pointwise description, i.e., the high-wavevector cutoff, by using a
renormalization procedure. For instance, one may choose a
``microscopic'' cutoff comparable with the inverse inclusion size, or a
much smaller cutoff corresponding to an elementary point size much
larger than the inclusion. The question then naturally arises of how
the constants of the pointwise quadratic potential will depend on the
coarse-graining scale.

Let us therefore consider a membrane described with an upper wavevector
cutoff $\Lambda=a^{-1}$. The membrane hosts a single pointwise inclusion
located at $\mathbf{r}=\mathbf{r}_1$. We shall perform a coarse-graining
of this system
in order that the new wavevector cutoff becomes $\Lambda'=(\mu
a)^{-1}<\Lambda$. We assume the whole system to be in thermal
equilibrium at the temperature $T$, however, since all energies are
quadratic, the renormalization will be athermal and we can set
$k_{\mathrm B}T=1$ without loss of generality. The partition function
can then be written as
\begin{equation}
\label{Z}
{\cal Z}=\int\!\!{\cal D} h\,\exp\left(-F_{\rm
m}\left[h\right]-U_1\left[h\right]\right),
\end{equation}
where the inclusion's pointwise potential can be written as
$U_1=\frac{1}{2}\Gamma(\mathsf K_1-\mathsf C_1)^{\mathsf t} {\mathsf
N}_1(\mathsf K_1-\mathsf C_1)$, as in Eq.~(\ref{u1plusu2}). Here,
$\mathsf N_1$ is the $3\times3$ uper-left block of the matrix $\mathsf
N$ of Eq.~(\ref{Nmatrix}), and
$\mathsf{K_1^t}=(H^{(1)}_{11},H^{(1)}_{12},H^{(1)}_{22})$,
$\mathsf{C_1^t}=(c,0,c)$. It is convenient to write $\exp(-U_1)$ as the
result of a Gaussian integral (Hubbard-Stratonovich
transformation~\cite{chaikinlubensky}):
\begin{equation}
\exp(-U_1)\propto\int\!\!d^3\Phi\,
\exp\left[-\frac{1}{2}\Phi^{\mathsf t}\mathsf N_1^{-1}\Phi
+i\Phi^{\mathsf t}\left(\mathsf C_1-\int\frac{d^2q}{(2\pi)^2}\mathsf {K}_1
({\bf q})e^{i\mathbf{q}\cdot\mathbf{r}_1}\right)\right],
\end{equation}
where 
${\mathsf{K}_1}^{\mathsf t}({\bf q})=\left(-q_1^2h({\bf
q}),-q_1q_2h({\bf q}),-q_2^2h({\bf q})\right)$.  Then, going to the
variables $h(\bf q)$, Eq.~({\ref{Z}}) becomes (up to a multiplicative
constant):
\begin{equation}
{\cal Z}=\int \!\!d^3\Phi\exp
\left(
-\frac{1}{2}\Phi^{\mathsf t}\mathsf N_1^{-1}\Phi
+i\Phi^{\mathsf t}\mathsf C_1
\right) 
\int\!\!\!\prod_{q<1/a}\!\!\!dh({\bf q})\,
\exp\left[
-\int\!\!\frac{d^2q}{(2\pi)^2}
\frac{\kappa}{2}
q^4|h(\mathbf{q})|^2+i\Phi^{\rm t}{\mathsf K}_1({\bf q})
e^{i\mathbf{q}\cdot\mathbf{r}_1}\right]
\end{equation}
Since the argument of the second exponential in the above equation is a
quadratic function of $h(\bf q)$, it is straightforward to integrate
over the $h(\bf q)$ with wavevectors in the range $(\mu
a)^{-1}<|\mathbf{q}|<a^{-1}$. Once this integration is performed, the
resulting integral over $\Phi$ is Gaussian, and yields, after some
elementary algebra, 
\begin{equation}
{\cal Z}=\int\!\!{\cal D}h^<\,\exp
\left(-F_m[h^<]-U_1^{\rm eff}[h^<]\right),
\end{equation}
where $h^<({\bf r})=(2\pi)^{-2}\int_{q<1/(\mu a)}\!d^2q\,h({\bf
q})e^{i\mathbf{q}\cdot\mathbf{r}}$, and
\begin{eqnarray}
U_1^{\rm eff}&=&\frac{\Gamma^{\rm eff}}{2}
\left(\partial_\alpha\partial_\beta h^<({\bf r}_1)
-C_{\alpha\beta}^{(1)}\right)^2
+\frac{\Gamma^{\rm eff}\epsilon^{\rm eff}}{2}\partial_\alpha\partial_\alpha h^<({\bf r}_1)\partial_\beta\partial_\beta h^<({\bf r}_1)
,\\
\Gamma^{\rm eff}&=&\Gamma\frac{\kappa\,b_0^2}{\kappa\,
b_0^2+2\Gamma\left(1-\mu^{-2}\right)},\\
\epsilon^{\rm eff}&=&\frac{\epsilon\kappa
b_0^2+\Gamma\left(1+2\epsilon\right)
\left(\mu^2-1\right)}
{\kappa\,b_0^2\mu^2-4\Gamma\left(1+2\epsilon\right)\left(\mu^2-1\right)},
\end{eqnarray}  
where we have defined $b_0=4\sqrt{2\pi}a$.  This set of exact equations
is our second main result. It describes the effective quadratic coupling
between the inclusion and the membrane in the coarse-grained pointwise
description corresponding to the new cutoff $(\mu a)^{-1}$. Note that
the inclusion's preferred curvature $\mathsf{C}_1$ is unaffected by the
coarse-graining procedure. This result may seem surprising, however one
should realize that the actual curvature set by the inclusion will
naturally be smaller in the coarse-grained description due to the fact
that $\Gamma^{\rm eff}<\Gamma$.

In principle, the interaction between two inclusions should be
independent of the level of coarse-graining chosen to describe the
inclusions (as long as their separation is larger than the length-scale
of the cutoff). One can indeed check that the interaction
Eq.~(\ref{fint}) is unchanged under the substitutions $b\equiv b_0\to\mu
b_0$, $\Gamma\to\Gamma^{\rm eff}$ and $\epsilon\to\epsilon^{\rm eff}$.

Now, since the rigidities $\Gamma$ and $\epsilon$ are scale-dependent,
an inclusion may be hard or soft depending on the coarse-graining level.
Indeed, according to the invariance of the interaction mentioned above,
the criterion for the hardness or softness of the inclusion is to
compare $\Gamma^\mathrm{eff}/\kappa$ to $(\mu b_0)^2$. For $\mu\gg1$,
the ``rigidity'' $\Gamma/(\kappa b_0^2)$, which becomes
$\Gamma^\mathrm{eff}/(\kappa\mu^2b_0^2)$ after coarse-graining, is
reduced by a factor $\simeq\!\mu^2[1+2\Gamma/(\kappa b_0^2)]\gg1$. Hence
all inclusions effectively become soft at large coarse-graining scales. 

Is this in contradiction with the fact that hard inclusions have an
interaction proportional to $\kappa$ while soft inclusions have an
interaction inversely proportional to $\kappa^2$? The answer is no,
because the effective constants $D$ and $A$ introduced by Misbah et
al.~\cite{misbah02} actually depend on $\kappa$. (This is
another way to see that there is no discrepancy between the results of
Ref.~\cite{misbah02} and those of the previous works.) Indeed, starting
from a hard inclusion ($\Gamma\gg\kappa b_0^2$), we obtain, in the limit
$\mu\gg1$ where all inclusions become soft,
$\Gamma^\mathrm{eff}\simeq\frac{1}{2}\kappa b_0^2$ and
$\epsilon^\mathrm{eff}\simeq-\frac{1}{4}$, while the preferred curvature
$c_0$ is unchanged. With the correspondance
$D^\mathrm{eff}=-\Gamma^\mathrm{eff}$ and
$A^\mathrm{eff}=c_0\Gamma^\mathrm{eff}$, we obtain
$(A^\mathrm{eff})^2D^\mathrm{eff}\propto\kappa^3$, hence
Eq.~(\ref{soft}) indeed yields an interaction proportional to $\kappa$.

We therefore conclude that the correct procedure to determine whether a
given inclusions is \textit{fundamentally} hard or soft is to estimate
$\Gamma$ at the coarse-graining level where the cutoff $a$ compares with
the inclusion size. The corresponding ``microscopic" value $\Gamma_m$
must then, according to the discussion at the end of Sec.~\ref{induced},
be compared with $\kappa a^2$ (since $b$ and $a$ are expected to be of
the same order of magnitude). An inclusion will be fundamentally hard if
$\Gamma_m\gg\kappa a^2$ or fundamentally soft if $\Gamma_m\ll\kappa
a^2$.

\section{Proteins as ``hard'' inclusion}

We now argue that integral proteins of conical shape should be
fundamentally considered as \textit{hard} inclusions. First, it is very
likely that the membrane elasticity is too weak to alter the conical
shape of the protein, because of the strong dipolar interactions within
the bundles of $\alpha$-helices in the transmembrane domain. Beside, a
significant change in the protein shape would most probably alter its
function. We can therefore assume that the protein behaves as a rigid
conical body and that the weakness of the curvature potential embodied
in the parameter $\Gamma_m$ arises from the \textit{tilt} degree of
freedom of the lipids. Let us consider a conical protein having a
circular section of radius $R$ in the membrane plane and an aperture
angle $2\theta$. An estimate of $\Gamma_m$ can be obtained by
calculating the energy stored in the lipid tilt when one assumes that
the membrane midplane remains perfectly flat. The latter is $W\approx
B\theta^2\times2\pi R\times\xi$, where $B$ is the elastic constant
associated with the tilt and $\xi\approx(\kappa/B)^{1/2}$ the tilt
relaxation length. With $c_0=\theta/R$, which corresponds to the
membrane curvature that the protein attempts to set, this energy can be
rewritten as $W\approx\frac{1}{2}\Gamma_m c_0^2$ with
$\Gamma_m\approx4\pi\kappa R^3/\xi$. Hence we obtain \begin{equation}
\frac{\Gamma_m}{\kappa a^2}\approx4\pi\frac{R^3}{a^2\xi}.
\end{equation} Interestingly, this quantity is independent of $\kappa$
and it grows as $R^3$. With typically $R\simeq6\,\mathrm{nm}$,
$a\simeq\xi\simeq3\,\mathrm{nm}$, we obtain $\Gamma_m/(\kappa
a^2)\approx100$, which justifies treating conical proteins as a hard
inclusion.

\section{Conclusion}

We have calculated the induced interaction acting between
pointwise membrane inclusions that locally interact up to quadratic
order with the membrane curvature tensor. According to the rigidity of
the potential constraining the local membrane curvature, inclusions can
be viewed as hard or soft.  The interaction between hard inclusions is
proportional to $\kappa$ while the interaction between soft inclusions
is inversely proportional to $\kappa^2$ (both are also inversely
proportional to the fourth power of the separation in the case of
isotropic inclusions). This reconciliates the recent view of Marchenko
and Misbah~\cite{misbah02} with those of earlier works.

To discriminate whether a given inclusion is fundamentally hard
or soft, one should however consider the coupling constants defined at
the microscopic level, i.e., defined in a pointwise description in which
the elementary point-size (the cutoff of the model) compares with the
inclusion size. Indeed, we have shown that coarse-graining the pointwise
description to length-scales much larger that the inclusion's size
transforms any hard inclusion into a soft one, however with coupling
constants depending on $\kappa$ in such a way that the global
interaction remains proportional to $\kappa$ if the inclusions was
fundamentally hard. In the light of this analysis we have shown that
transmembrane proteins should be considered as fundamentally hard
inclusions.

\begin{acknowledgments}
Helpful interactions with Paolo Galatola are gratefully acknowledged.
\end{acknowledgments}


\end{document}